# Quantized anomalous Hall resistivity achieved in molecular beam epitaxy-grown MnBi$_2$Te$_4$ thin films

Yunhe Bai[1]*, Yuanzhao Li[1]*, Jianli Luan[1]*, Ruixuan Liu[1]*, Wenyu Song[1], Yang Chen[1], Peng-Fei Ji[1], Qinghua Zhang[7], Fanqi Meng[6], Bingbing Tong[3], Lin Li[3], Yuying Jiang[1], Zongwei Gao[3], Lin Gu[6], Jinsong Zhang[1,2,5], Yayu Wang[1,2,5], Qi-Kun Xue[1,2,3,4], Ke He[1,2,3,5]†, Yang Feng[3]†, and Xiao Feng[1,2,3,5]†

[1]*State Key Laboratory of Low Dimensional Quantum Physics, Department of Physics, Tsinghua University, Beijing 100084, China*
[2]*Frontier Science Center for Quantum Information, Beijing 100084, China*
[3]*Beijing Academy of Quantum Information Sciences, Beijing 100193, China*
[4]*Southern University of Science and Technology, Shenzhen 518055, China*
[5]*Hefei National Laboratory, Hefei 230088, China*
[6]*School of Materials Science and Engineering, Tsinghua University, Beijing 100084, China*
[7]*Institute of Physics, Chinese Academy of Sciences, Beijing 100190, China*

* *These authors contributed equally to this work.*

† *Corresponding author. Email:* kehe@tsinghua.edu.cn (K. H.); fengyang@baqis.ac.cn (Y. F.); xiaofeng@mail.tsinghua.edu.cn (X. F.)

## Abstract

The intrinsic magnetic topological insulator MnBi$_2$Te$_4$ provides a feasible pathway to high temperature quantum anomalous Hall (QAH) effect as well as various novel topological quantum phases. Although quantized transport properties have been observed in exfoliated MnBi$_2$Te$_4$ thin flakes, it remains a big challenge to achieve molecular beam epitaxy (MBE)-grown MnBi$_2$Te$_4$ thin films even close to the quantized regime. In this work, we report the realization of quantized anomalous Hall resistivity in MBE-grown MnBi$_2$Te$_4$ thin films with the chemical potential tuned by both controlled in-situ oxygen exposure and top gating. We find that elongated post-annealing obviously elevates the temperature to achieve quantization of the Hall resistivity, but also increases the residual longitudinal resistivity, indicating a picture of high-quality QAH puddles weakly coupled by tunnel barriers. These results help to clarify the puzzles in previous experimental studies on MnBi$_2$Te$_4$ and to find a way out of the big difficulty in obtaining MnBi$_2$Te$_4$ samples showing quantized transport properties.

**Introduction**

The quantum anomalous Hall (QAH) effect is a quantum Hall (QH) effect induced by spontaneous magnetization of a material and thus free from external magnetic field [1]. After the first experimental realization in molecular beam epitaxy (MBE)-grown thin films of a magnetically doped topological insulator (TI) in 2013 [2], the QAH effect has attracted rapidly growing research interests as a route towards various other exotic quantum phases such as axion insulator and chiral topological superconductor [3,4]. Magnetically doped TIs need a rather low temperature (usually below 1 K) to exhibit the QAH effect and are highly disordered due to the randomly distributed magnetic dopants [5,6]. In recent years, two kinds of intrinsic QAH systems have been discovered, namely, thin films/flakes of $MnBi_2Te_4$-family compounds [7–11] and moiré superlattices of two-dimensional (2D) materials [12–14], providing potentially cleaner and more robust alternatives to magnetically doped TIs. Particularly, the huge magnetic gap predicted in $MnBi_2Te_4$ films (~50 meV) heralds a high temperature QAH effect. Nearly quantized transport properties have indeed been observed in $MnBi_2Te_4$ thin flake samples up to ~40 K [15]. However, $MnBi_2Te_4$ thin flakes (also moiré superlattices) rely on rather challenging and tricky fabrication techniques, and suffer from the irregular shape, small size, and ill-controlled performance of the samples. Actually, only a few groups world-wide have ever succeeded in obtaining $MnBi_2Te_4$ flakes showing quantized transport properties at a rather low yield even with the best single crystals. The intractable thin flake samples make systematic studies and optimizations very difficult, leading to seemingly inconsistent data from different samples and measurements [11,15–20]. Moreover, to explore QAH-based novel topological quantum states [21,22] and their electronic applications, $MnBi_2Te_4$ samples should be prepared in a repeatable, controllable and scalable way so that they can construct complex and large-area heterostructures, arrays and circuits for different purposes. It is very difficult for thin flake samples to satisfy the requirement.

Molecular beam epitaxy can provide large, regular and well-controlled samples with excellent repeatability and tunability, particularly advantageous in scalable preparation of QAH-based heterostructures. Although MBE growth of MnBi$_2$Te$_4$ thin films has long since been achieved, the largest Hall resistivity obtained in those samples merely reached half of the quantized value ($h/e^2$) so far [10,23–27]. In this work, we achieved MBE-grown MnBi$_2$Te$_4$ thin films showing quantized Hall resistivity in the ferromagnetic (FM) configuration under high magnetic field. The much larger sample size and higher repeatability of sample properties enable us to systematically study their properties and clarify some puzzles in previous works on MnBi$_2$Te$_4$.

**Results and discussion**

The schematic procedure of sample preparation is shown in Fig. 1a. MnBi$_2$Te$_4$ films are grown on sapphire (0001) substrates by co-evaporation of Mn, Bi and Te and post-annealing at 270°C (see Methods, SI A and ref. [28] for the detailed MBE growth procedure and parameters). Figure 1b shows the reflection high energy electron diffraction (RHEED) pattern of a MnBi$_2$Te$_4$ film (the inset is that for the sapphire substrate). The as-grown MnBi$_2$Te$_4$ films are electron-doped due to unavoidable occupying of Bi atoms on Mn sites [29], similar to single crystal samples. We apply in-situ oxygen exposure and electric gating to coarsely and finely tune the Fermi level into the surface state gap, respectively [30]. The as-grown films are transferred to another chamber of the same MBE system and exposed to ~1×10$^{-3}$ mbar O$_2$ for an hour at room temperature. The pressure resembles the O$_2$ partial pressure in a glove box for exfoliating MnBi$_2$Te$_4$ flakes. The oxygen pressure is slightly modified for samples of different thicknesses to keep them roughly charge neutral. Such a medium O$_2$ exposure effectively reduces the carrier density of MnBi$_2$Te$_4$ films with little influence to their electronic structures and properties (Figs. S2 and S3). After O$_2$ exposure, an 8-nm-thick cadmium selenide (CdSe) capping layer is deposited onto the films to avoid uncontrolled charge doping in the following fabrications and measurements. The atomic force microscopy and high-resolution scanning transmission electron microscopy (STEM) images of a nominally five-septuple-layer (5-SL) MnBi$_2$Te$_4$ sample with CdSe capping layer are shown in Figs. 1c and 1d, respectively. The film is mainly composed of one terrace of uniform thickness, distributed with small islands and depressions of 1-SL height.

The MnBi$_2$Te$_4$ films are lithographed by argon ion milling into Hall bars (40 μm × 80 μm) through a molybdenum mask. An AlO$_x$ dielectric layer is then grown on the samples by atomic layer deposition, and the top-gate electrode is deposited on it through another molybdenum mask (the upper limit of the sample size is determined by the quality of the dielectric layer). The false-color photo of a final device is displayed in Fig. 1e. The sample temperature is kept below 60°C during the whole top-gate structure fabrication procedure to avoid sample degradation. The low temperature tolerance may come from the oxygen introduced in the films. Different devices on a 3 mm × 5 mm sample show similar transport properties (Fig. S9), indicating the excellent homogeneity and repeatability of the MBE-grown MnBi$_2$Te$_4$ samples. Figure 1f displays the longitudinal resistivity ($\rho_{xx}$)–temperature ($T$) curve of a 5-SL MnBi$_2$Te$_4$ thin film (with the gate voltage $V_g$ = 0 V), which exhibits a clear cusp at 21.8 K corresponding to the Néel temperature ($T_N$) and an insulating behavior at lower temperature [31,32].

Figures 2a and 2b show the magnetic field ($\mu_0H$) dependences of the Hall ($\rho_{yx}$) and longitudinal ($\rho_{xx}$) resistivities, respectively, of a 5-SL sample (at charge-neutral point) measured at five different temperatures. Figure 2c displays the $\rho_{yx}$–$V_g$ and $\rho_{xx}$–$V_g$ curves at $T$ = 0.02 K and $\mu_0H$ = -9 T. The sample was post-annealed at 270°C for 30 min. after MBE growth (referred below as 5 SL-1). The $\rho_{yx}$–$\mu_0H$ and $\rho_{xx}$–$\mu_0H$ curves have the typical shapes of MnBi$_2$Te$_4$: the high field (> 6.5 T) and low field (< 3 T) parts of the curves roughly correspond to MnBi$_2$Te$_4$ in the FM and antiferromagnetic (AFM) configurations, respectively. In the FM state around ±9 T, |$\rho_{yx}$| and $\rho_{xx}$ always evolve inversely, with magnetic field, gate voltage and temperature, which is a signature of the dissipationless quantum Hall or QAH edge state. The former one is excluded by the absent transport evidence of Landau levels, such as Shubnikov de Haas (SdH) peaks. Therefore, the QAH state dominates the high-field transport properties. At 0.02 K, |$\rho_{yx}$| reaches 0.98$h/e^2$, and $\rho_{xx}$ drops to 0.20$h/e^2$ at ±9 T, exhibiting a decent quantization. Descendible deviation of $\rho_{yx}$ from quantization is observed at 0.4 K (0.93$h/e^2$). It is difficult to accurately estimate the activation gap from the $\sigma_{xx}$–$T^{-1}$ curve (Fig. S14). But according to Figs. 2a and 2b, the gap size should be at the same level as that of Cr-doped $(Bi,Sb)_2Te_3$ films, much lower than that of good $MnBi_2Te_4$ flake samples. In the AFM state at low field (between ±3 T), the $\rho_{yx}$–$\mu_0H$ curves exhibit a hysteresis loop, with the coercive field ($H_c$) of ~1.1 T and zero field $\rho_{yx}$ of ~$0.34h/e^2$ at 0.02 K (Fig. 2a). $\rho_{xx}$ at high field decreases with decreasing temperature, as expected in a QAH system, meanwhile at low field, $\rho_{xx}$ increases with decreasing temperature, showing an insulating behavior (Fig. 2b).

We systematically investigate thickness-dependent transport properties of the $MnBi_2Te_4$ films from 2 SL to 7 SL, all post-annealed at 270°C for 30 min, as shown in Fig. S5. The 4-SL to 6-SL samples all show a nearly quantized $\rho_{yx}$ and a $\rho_{xx}$ minimum at high field in both the $\mu_0H$- and $V_g$-dependent curves, indicating the QAH states in the FM state. No signatures of the QAH state are observed in the 2-SL and 3-SL films. The absence of the QAH state in the 3-SL film, which ought to be in the QAH regime according to early works [7,11], may be attributed to the influence of oxidation [33]. In the 7-SL film, the signatures of the QAH state can hardly been distinguished due to parallel conduction channels.

We found that elongated post-annealing significantly changes the transport properties of the films. Figures 2d and 2e display the $\rho_{yx}$–$\mu_0H$ and $\rho_{xx}$–$\mu_0H$ curves, respectively, of a 5-SL $MnBi_2Te_4$ film with post-annealing at 270°C for 120 min. (referred as 5 SL-2) measured at different temperatures. The annealing was carried out in Te flux to avoid Te desorption. Notably, $|\rho_{yx}|$ of the sample reaches full quantization ($0.99h/e^2$) at 0.02 K at high field (the strong noise in the $\rho_{yx}$–$\mu_0H$ curve between ±5 T is due to the huge $\rho_{xx}$, which will be discussed below). The Hall resistivity keeps $0.98h/e^2$ at 0.4 K, $0.97h/e^2$ at 0.7 K, and still $0.89h/e^2$ even at 2 K. Obviously, the $\rho_{yx}$ quantization at high field becomes more robust to temperature, comparable to some $MnBi_2Te_4$ thin flake samples. On the other hand, $\rho_{xx}$ is significantly enhanced, rather than reduced as expected in a good QAH sample, by the elongated annealing. The high-field $\rho_{xx}$ of the sample is $0.73h/e^2$ at ±9 T, 0.02 K, and rises with increasing temperature, suggesting that the ground state is still the QAH phase. The low field $\rho_{xx}$ at 0.02 K becomes extremely large (the measured $\rho_{xx}$ of ~$200h/e^2$ is not accurate due to the too large resistance), and drops rapidly with increasing temperature (about $5h/e^2$ even at 2 K), exhibiting a typical insulating behavior. It suggests absence of dissipationless chiral edge states in the sample at low field, which will be discussed later.

The effect of annealing duration can be clearly seen in the $\rho_{yx}$–$V_g$ and $\rho_{xx}$–$V_g$ curves measured at -9 T and 0.02 K of the 5 SL-1 (Fig. 2c) and 5 SL-2 (Fig. 2f) samples, and also in their temperature dependences of $\rho_{yx}$ (Fig. 3a) and $\rho_{xx}$ (Fig. 3b) at -9 T. Obviously, elongated annealing drives $\rho_{yx}$ closer to the quantized value ($h/e^2$) but makes $\rho_{xx}$ farther away from zero at high field. Longer post-annealing duration time (at 270°C) does not obviously change the sample properties further, as shown in Fig. 3c. So 120 min. post-annealing is selected representatively. Similar annealing effect is also observed in the films of other thicknesses (Fig. S14).

In a well quantized QAH film, the dissipationless chiral edge channel surrounding the whole sample contributes to quantized $\rho_{yx}$ and vanishing $\rho_{xx}$ (see Fig. 3e). With dissipative conduction channels of residual bulk carriers, $\rho_{xx}$ becomes non-zero, and $\rho_{yx}$ drops below the quantized value (Fig. 3f). The early results on the QAH effect and the data of the $MnBi_2Te_4$ film with 30 min. annealing (Figs. 2a and 2b) are all consistent with this case. It thus looks confusing that the sample with 120 min. annealing has a perfectly quantized $\rho_{yx}$ below 0.4 K but, at the same time, keeps a large $\rho_{xx}$ at high field (Figs. 2d and 2e). The quantized $\rho_{yx}$ accompanied by a significant $\rho_{xx}$ was ever observed and discussed in two-dimensional electron systems, known as the "quantized Hall insulator", which is composed of weakly coupled quantum Hall regions (puddles) [34], each surrounded by dissipationless quantum Hall edge states. The tunneling between the neighboring quantum Hall puddles contributes to a large $\rho_{xx}$ of the whole sample which is determined by the transmission probability (Fig. 3g) [35], meanwhile $\rho_{yx}$ can still remain the quantized value. Although the $MnBi_2Te_4$ films with elongated annealing have a QAH ground state at high field, not really an insulator, the high residual $\rho_{xx}$ can be understood with this picture. Namely, the samples are composed of well-quantized

QAH puddles with the tunneling between the puddles giving a large $\rho_{xx}$ of the whole sample.

Through ordinary Hall effect measurements above $T_N$, we found that the MnBi$_2$Te$_4$ films with longer-time post-annealing have a lower carrier density at the charge-neutral point (Fig. 3d). Probably, long-time annealing aggregates the defects or impurities that induce the chemical potential fluctuation in a MnBi$_2$Te$_4$ film [36], e.g. Mn atoms occupying Bi sites. As a result, most areas of the film become cleaner with less chemical potential fluctuations, acting as high-quality QAH puddles. But at their boundaries where defects aggregate, the QAH phases are destroyed, and the dissipationless edge states of QAH puddles can only tunnel through there, contributing to the non-zero $\rho_{xx}$. Therefore, the residual $\rho_{xx}$ is expected to be reduced in MnBi$_2$Te$_4$ films of smaller size which includes fewer QAH puddles. Our preliminary result indeed shows that a MnBi$_2$Te$_4$ device of smaller size exhibits a lower $\rho_{xx}$ under high field than a device of larger size (Fig. S12), while a systematic size-dependent transport study is needed to confirm it. Similar situation may also happen in MnBi$_2$Te$_4$ single crystals which usually experience a rather long-time annealing. Consequently, full quantization is observed in exfoliated thin flakes (usually several μm large) only including clean areas, but not in those with aggregated defects. It explains the low yield of MnBi$_2$Te$_4$ thin flake samples showing perfect quantized transport properties.

Figures 4a and 4b display the $\mu_0 H$ dependences of the longitudinal ($\sigma_{xx}$) and Hall ($\sigma_{xy}$) conductivities measured at different temperatures of the 5-SL films with 30 min. and 120 min. post-annealing, respectively. At high field, the former sample has nearly quantized $\sigma_{xy}$, meanwhile the latter one has $\sigma_{xy}$ obviously lower than $e^2/h$ due to the larger $\rho_{xx}$, despite the fully quantized $\rho_{yx}$. The ground states of the films can be checked by the tendency of the conductivity tensor ($\sigma_{xy}$, $\sigma_{xx}$) evolution with increasing magnetic field and decreasing temperature (the flow charts), as shown in Fig. 4c. Here, the data of 5-SL, 6-SL and 7-SL films are plotted, and the left and right parts correspond to the data from the long- and short-time post-annealed samples, respectively. Similar to early results of the quantum Hall and QAH effects [37–40], the conductivity tensors tend to flow towards (0, 0) or ($e^2/h$, 0) points at low temperature. The (0, 0) point is the insulating phase (also known as the Hall insulator), and the ($e^2/h$, 0) point is the QAH or quantum Hall phases [41–44].

In the FM configuration (high field), the conductivity tensors of the films all flow towards the ($e^2/h$, 0) points with increasing magnetic field (indicated by the black arrow) and decreasing temperature, which indicates a QAH ground state (the quantum Hall phase is excluded as discussed above). In the AFM configuration (low field), the conductivity tensors all flow to the (0, 0) point with decreasing magnetic field (the opposite direction of the black arrow) and temperature, regardless of odd- or even-SL films, which indicates an insulating ground state as directly observed in Figs. 2b and 2e. The $\rho_{xx}$–$\mu_0 H$ curves of the two 5-SL samples measured at different temperatures all cross at one point, signifying a quantum phase transition from the QAH state (high field) to the insulating state (low field). A scaling analysis near the critical magnetic field ($H_{cr}$) is shown in the insets of Figs. 4d and 4e, where the relation between $\rho_{xx}$ and the scaling variable ($H$-$H_{cr}$)/$T^\kappa$ are plotted. $\rho_{xx}$ collapses into two branches contributed by two sides of $H_{cr}$, respectively, with the fitted scaling exponent $\kappa$ = 0.45, identical to that for the phase transitions in QH systems [45].

The insulating behavior in the odd-SL films at low field looks inconsistent with the prediction that MnBi$_2$Te$_4$ in the AFM configuration is a Chern insulator in the odd-SL films [7–9]. One possible cause is the thickness fluctuations of films. The 1-SL islands and depressions as shown in Fig. 1c make an odd-SL QAH film mixed with even-SL axion insulator regions, which could elevate $\rho_{xx}$. However, according to early theoretical and experimental results on magnetically modulation-doped TI films [46], even when half of a QAH film becomes axion insulator, $\rho_{xx}$ only increases to ~1$h/e^2$ which is much lower than the observed $\rho_{xx}$ of up to hundreds $h/e^2$ in the MnBi$_2$Te$_4$ films with 120 min. annealing. Actually, we found that the measured $\rho_{xx}$ of odd-SL films is not sensitive to the coverage of the even-SL regions (Fig. S13). The zero-magnetic-field insulating state has also been reported in many odd-SL MnBi$_2$Te$_4$ flakes that show quantized $\rho_{yx}$ at high field, without thickness fluctuation [15,17,19,47]. The conductivity tensor data of MnBi$_2$Te$_4$ thin flakes at zero field around 1.5 K are also plotted in Fig. 4c for comparison. Most of the data are near the (0, 0) point, suggesting an insulating ground state. The insulating ground state is confirmed by the huge resistivity observed in our MBE films. Only in very rare odd-SL flake samples, has the QAH effect at zero magnetic field been observed [11].

Another possibility is that many AFM domains exist in the films around zero magnetic field, instead of the single AFM domain case assumed by the theoretical work predicting the QAH effect. Antiferromagnetic domains have been observed in MnBi$_2$Te$_4$ single crystals by magnetic force microscopy with the typical size of ~10 μm [48,49]. Since the domain size is much larger than the width of the QAH edge states in MnBi$_2$Te$_4$ (~1 μm) [20], the sample can be considered as a network of chiral edge states, which usually gives a $\rho_{xx}$ of the same order of magnitude as $h/e^2$ [50,51]. Again, the huge $\rho_{xx}$ observed in the samples with 120 min. annealing suggests absence of the chiral edge state network. However, the magnetic structure of the MnBi$_2$Te$_4$ thin films/flakes might be different from single crystals due to the low thickness. There are other possibilities, e.g. the influences of antisite defects [52], that might also lead to the insulating behavior in MnBi$_2$Te$_4$ thin films/flakes at zero field. The present data are not enough to decide which mechanism dominates. Nevertheless, the huge $\rho_{xx}$ at zero magnetic field and its sensitivity to the post-annealing condition may help to understand the large fluctuations in sample properties of MnBi$_2$Te$_4$ thin flakes. Besides, considering that the QAH state is probably absent in most MnBi$_2$Te$_4$ samples at zero magnetic field, whatever the reason is, it is reasonable that previous angle-resolved photoemission spectroscopy (ARPES) measurements on MnBi$_2$Te$_4$ (on large samples at zero magnetic field) failed to clearly demonstrate the magnetic gap opening as predicted theoretically [22].

**Conclusion**

In summary, we have achieved MBE-grown MnBi$_2$Te$_4$ thin films exhibiting quantized Hall resistivity at high magnetic field. Both even- and odd-SL films show insulating behavior around zero magnetic field. Elongating the post-annealing time is found to significantly enhance the temperature to reach $\rho_{yx}$ quantization (at high field), but at the same time increase the residual $\rho_{xx}$, indicating formation of QAH puddles with their chiral edge states tunneled with each other. The information obtained from the MBE-grown films with good repeatability may help to clarify the complexity in the experimental results on MnBi$_2$Te$_4$ and improve the material for explorations to high temperature QAH effect and other novel topological quantum effects.

## Methods

### Thin film growth

MnBi$_2$Te$_4$ thin films are grown on treated sapphire (0001) substrates in an ultrahigh-vacuum MBE system with base pressure better than $2.0\times10^{-10}$ mbar. The commercial substrates are annealed in a tube furnace at 1100°C for 3 hours under O$_2$ atmosphere to obtain flat terraces on the surfaces. The treated substrates are degassed at 400°C for 30 minutes before growth. High purity Mn (99.9998%), Bi (99.9999%) and Te (99.9999%) are co-evaporated with commercial Knudsen cells. Post-annealing process at growth temperature for 0.5~10 hours is implemented to further improve sample quality. (If the annealing time is longer than 2 hours, Te flux is applied to avoid Te desorption.) After growth, the samples are in-situ exposed to oxygen with different pressures for different thicknesses and then an 8-nm CdSe layer is capped on the top at room temperature. Before the following fabrication, the topography is scanned by atomic force microscopy (Bruker, Innova). See more details in SI A.

### Fabrication

The MnBi$_2$Te$_4$ films are lithographed by argon ion milling into Hall bars (40 μm × 80 μm) through a molybdenum mask. Then, a 40-nm AlO$_x$ layer is grown on the samples as a gate dielectric by atomic layer deposition. The top-gate electrode of 5 nm Ti/ 20 nm Au is deposited on it through another molybdenum mask. Indium is applied as Hall bar electrodes.

### Transport measurements

Magneto-transport measurements are performed in a commercial dilution fridge Oxford Instrument Triton 400 with a base temperature below 20 mK. Due to the large resistance at low temperature, we use delta mode, combining source meter and voltmeter. The DC current is applied by the digital source meter Keithley 6221. The voltage drop is measured by the voltmeter Keithley 2182A. At low temperature, the delay time between the start of DC current applied and voltage reading is from 1.6 s to 5.1 s for different samples. The longitudinal and Hall voltage drops $V_{xx}$ and $V_{yx}$ are detected separately in different magnetic field sweeps. For samples 5 SL-1, 6 SL-1 and 7 SL-1, the source-drain currents used in longitudinal ($I_{xx}$) and Hall ($I_{yx}$) measurements are both 1 nA. For samples 4 SL, 5 SL-2, 6 SL-2 and 7 SL-2, $I_{xx}$ and $I_{yx}$ are 1 nA and 10 nA, respectively. The current remains unchanged during the variation of temperature and gating. The top gate (AlO$_x$ dielectric) voltages are applied by a Keithley 2400 multimeter.

## Acknowledgements

We thank C. Liu for helpful discussions. We thank Q. Liu for his help with device fabrications. We also thank Z. Li and C. Guo for their help with ToF-SIMS measurements.

## Funding

This work was supported by the National Natural Science Foundation of China (92065206, 11904053), the National Key Research and Development Program of China (2018YFA0307100, 2017YFA0303303), the Innovation Program for Quantum Science and Technology (2021ZD0302502).

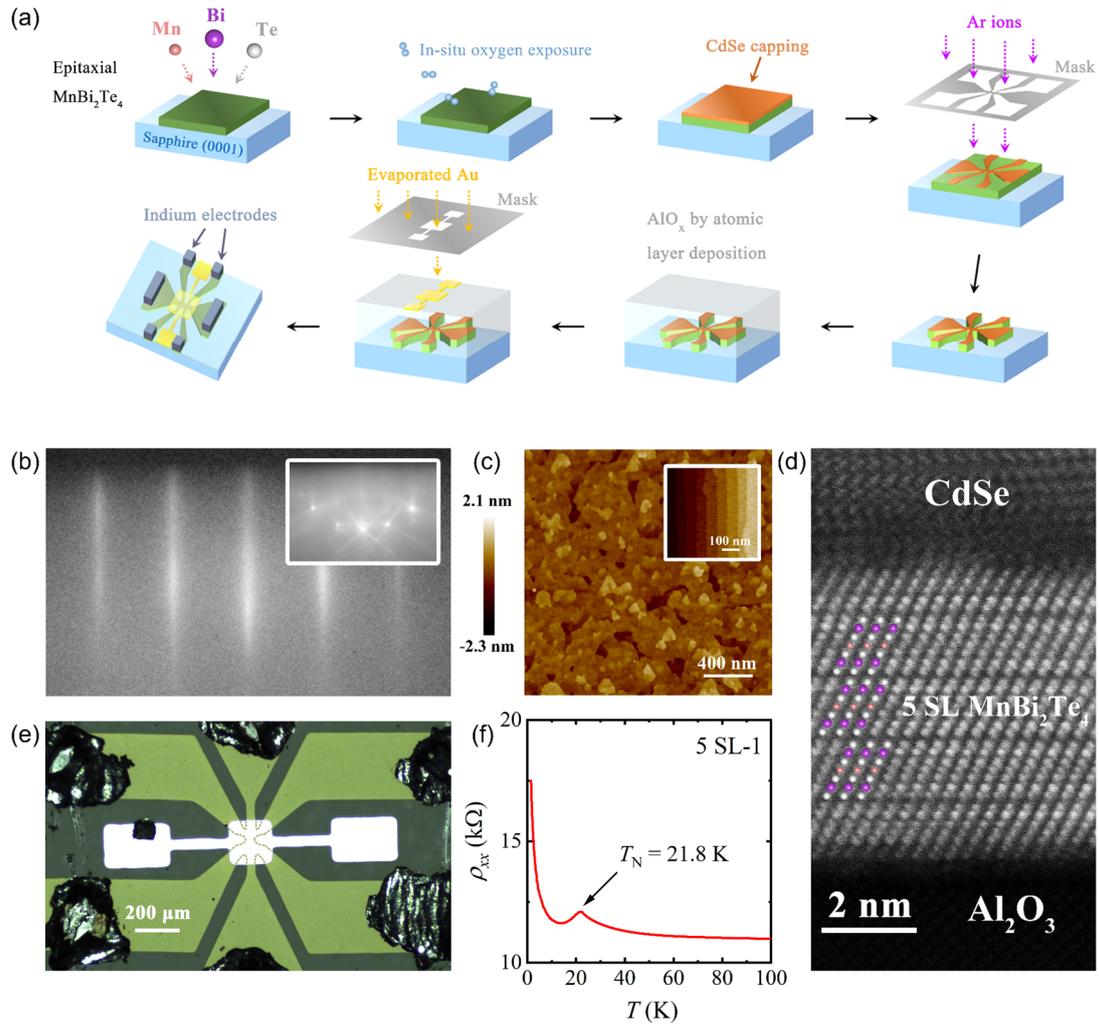

**Figure 1.** Characterizations of MnBi$_2$Te$_4$ thin films. (a) The schematic procedure of sample preparation. (b) RHEED patterns of an as-grown sample and its substrate (inset) in the same direction. (c) Topographies of sample 5 SL-1 with capping layer and a treated substrate (inset). (d) A typical STEM cross-sectional image of a 5-SL sample with same atom colors in (a). (e) Optical microscopy image of a typical device with the Hall bar size of 40 μm × 80 μm. The dashed line is guide to eye. (f) The $\rho_{xx}$–$T$ curve of sample 5 SL-1 with $V_g$ = 0 V.

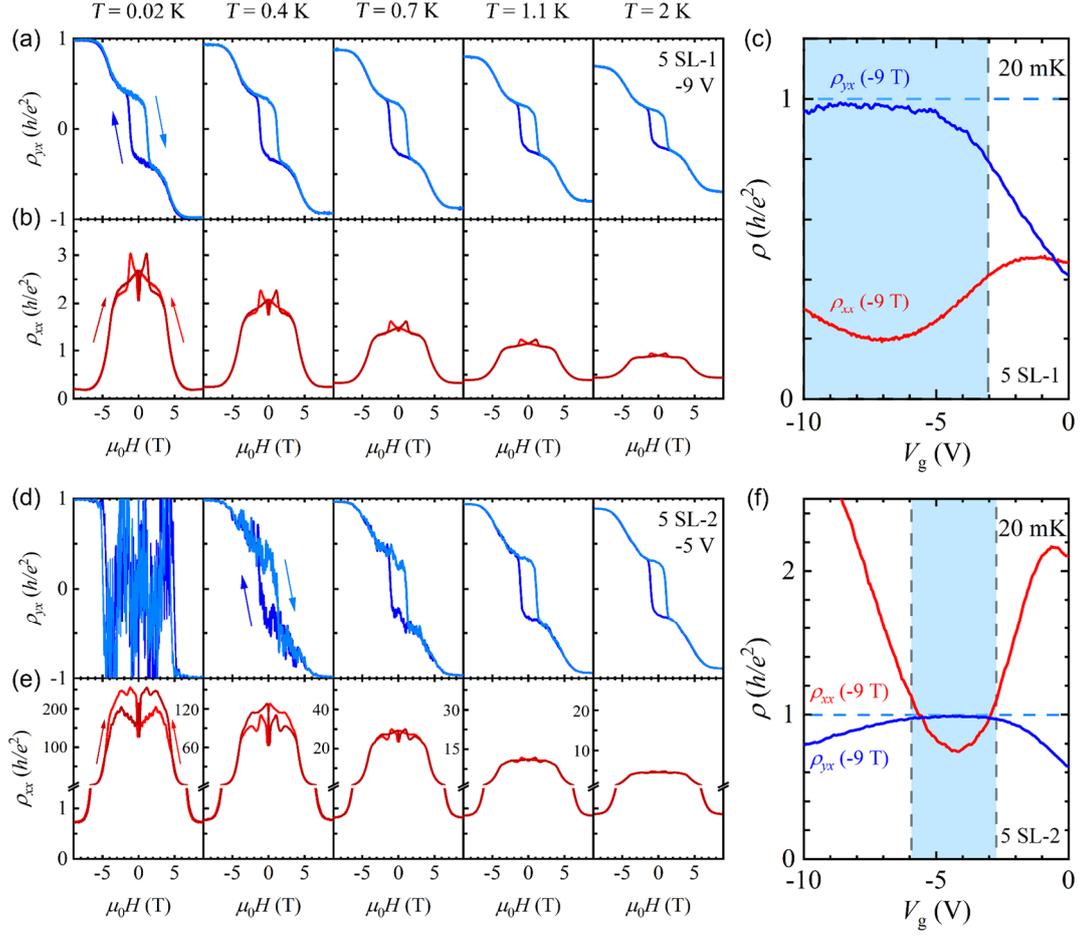

**Figure 2.** Magneto-transport measurements of 5-SL samples with different post-annealing time. (a, b) The $\rho_{yx}$–$\mu_0H$, $\rho_{xx}$–$\mu_0H$ curves of sample 5 SL-1 (post-annealed for 30 min.) at selected temperatures below 2 K, respectively. (c) The anti-symmetrized $\rho_{yx}$–$V_g$ and symmetrized $\rho_{xx}$–$V_g$ curves of sample 5 SL-1 under -9 T at 20 mK. The colored region represents the area with inverse $V_g$ dependences of $\rho_{yx}$ and $\rho_{xx}$. (d–f) The transport results of sample 5 SL-2 (post-annealed for 120 min.) with the same measurements as sample 5 SL-1 [corresponding to (a–c) respectively].

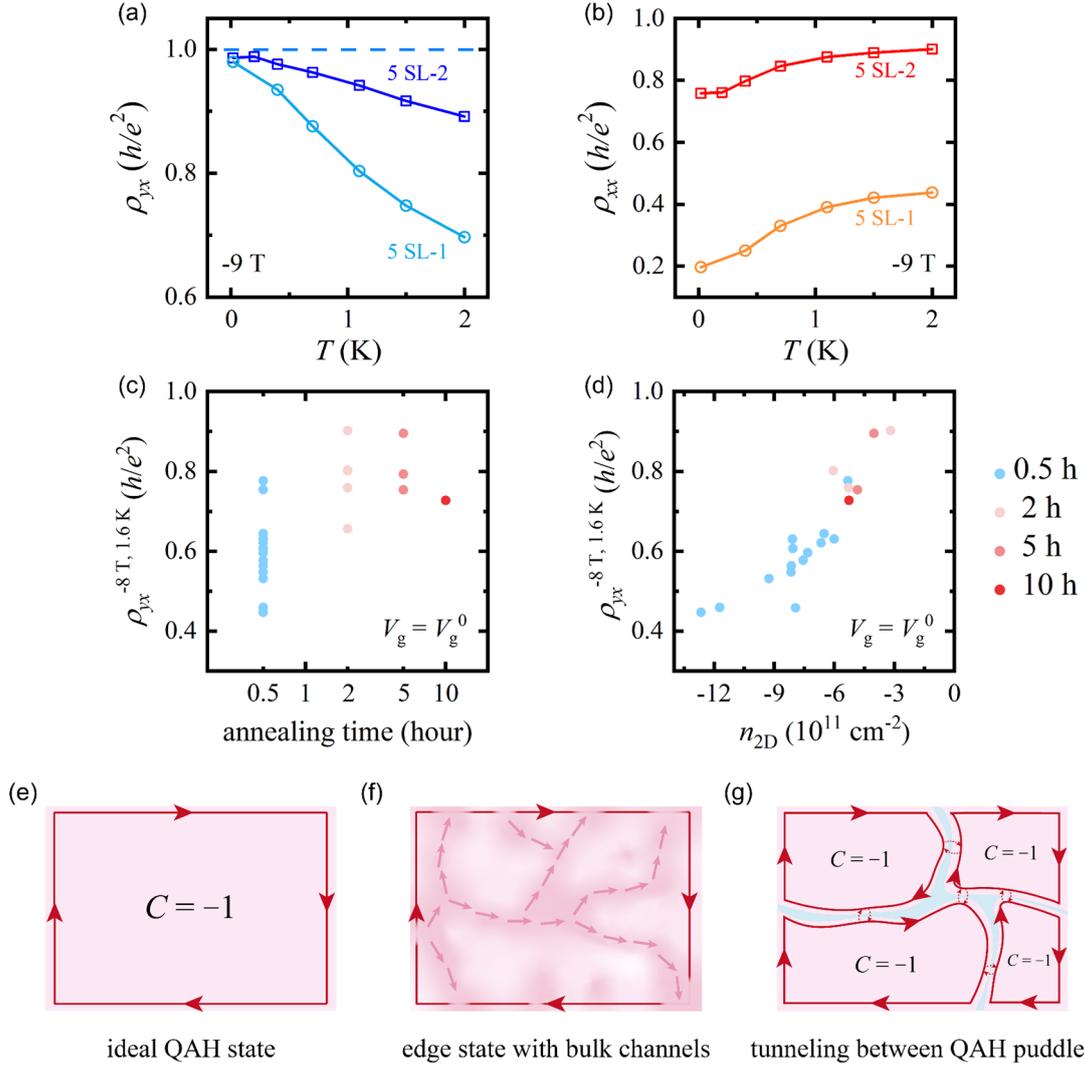

**Figure 3.** Transport properties of films with different post-annealing time in the ferromagnetic configuration. (a, b) The comparisons of $\rho_{yx}$–$T$, $\rho_{xx}$–$T$ curves of samples 5 SL-1 and 5 SL-2 under -9 T at the charge-neutral point, respectively. (c, d) The relations between $\rho_{yx}$ (-8 T) at 1.6 K and post-annealing time, $\rho_{yx}$ (-8 T) at 1.6 K and carrier density ($n_{2D}$), respectively. All the data are taken at the charge-neutral point ($V_g = V_g^0$) of each sample. (e–g) The schematic of three different scenarios of QAH state in realistic samples. Pink, purple and cyan regions represent QAH insulators, conductive bulk channels (with purple arrows) and insulating regions, respectively. Red solid and dashed lines represent edge states and tunneling processes between edge states, respectively.

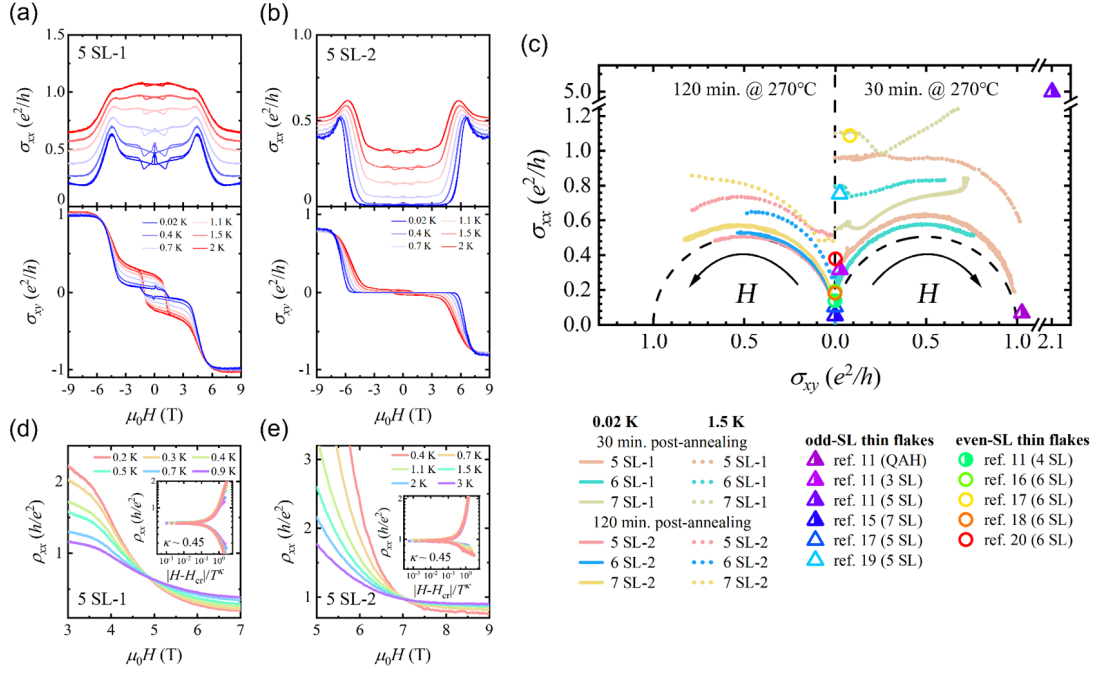

**Figure 4.** Quantum phase transitions in MnBi$_2$Te$_4$ thin films. (a, b) The temperature evolutions of $\sigma_{xx}$–$\mu_0H$ and $\sigma_{xy}$–$\mu_0H$ curves of sample 5 SL-1, sample 5 SL-2, respectively. (c) Phase diagram depicted by the conductivity tensor. MBE-grown thin films are labeled with different colors. The solid and dotted lines represent the data taken at the base temperature and 1.5 K, respectively. The left and right parts correspond to the data from the long- and short-time post-annealed samples, respectively. Both sides are $\sigma$ evolutions with increasing magnetic field (indicated by the black arrow). The triangles and circles represent conductivity tensor data of odd- and even-SL exfoliated thin flakes in previous reports around 1.5 K at 0 T, respectively. The aspect ratio of length ($L$) to width ($W$) of each device is used to calculate conductivities when they are shown (hollow symbols). If not, $L/W = 1$ is used (symbols colored half). The black dashed semicircle is guide to eye. (d, e) The quantum phase transitions near $H_{cr}$s of sample 5 SL-1, sample 5 SL-2, respectively. The insets show the scaling analyses.